\documentclass[a4paper]{article}
\usepackage{pdfsync}
\usepackage[utf8]{inputenc}
\usepackage{graphicx}
\usepackage{amssymb,amsfonts,amsmath,amsthm}
\usepackage{hyperref}
\usepackage{multirow}
\usepackage{verbatim}
\usepackage{xcolor}
\usepackage{booktabs}
\usepackage[sectionbib,round]{natbib}
\usepackage{textcomp}
\usepackage{upquote}
\usepackage{bm}
\usepackage{dirtree}
\usepackage{algorithm}%
\usepackage{algorithmicx}%
\usepackage{algpseudocode}%

\theoremstyle{plain}

\theoremstyle{definition}
\newtheorem{definition}{Definition}[section]

\newtheorem{remark}{Remark}[section]

\begin{document}

\title{\textbf{A multi-criteria decision support system to evaluate the effectiveness of training courses on citizens' employability}}


\author{Mar\'{\i}a C. Bas$^1$, Vicente J. Bol\'os$^1$, \'Alvaro E. Prieto$^2$, \\ Roberto Rodr\'{\i}guez-Echeverr\'{\i}a$^2$, Fernando Sánchez-Figueroa$^2$ \\ \\
{\small $^1$ Dpto. Matem\'aticas para la Econom\'{\i}a y la Empresa, Facultad de Econom\'{\i}a.} \\
{\small Universitat de Val\`encia. Avda. Tarongers s/n, 46022 Valencia, Spain.} \\
{\small $^2$ Dpto. Ingenier\'{\i}a Sistemas Inform\'aticos y Telem\'aticos.} \\
{\small Universidad de Extremadura. Avda. de la Universidad, 10071 C\'aceres, Spain.} \\
{\small e-mail\textup{: \texttt{maria.c.bas@uv.es}, \texttt{vicente.bolos@uv.es}, \texttt{aeprieto@unex.es}, }} \\
{\small \texttt{rre@unex.es}, \texttt{fernando@unex.es}}
}

\date{September 2024}

\maketitle

\begin{abstract}
 This study examines the impact of lifelong learning on the professional lives of employed and unemployed individuals. Lifelong learning is a crucial factor in securing employment or enhancing one's existing career prospects. To achieve this objective, this study proposes the implementation of a multi-criteria decision support system for the evaluation of training courses in accordance with their capacity to enhance the employability of the students. The methodology is delineated in four stages. Firstly, a `working life curve' was defined to provide a quantitative description of an individual's working life. Secondly, an analysis based on K-medoids clustering defined a control group for each individual for comparison. Thirdly, the performance of a course according to each of the four predefined criteria was calculated using a t-test to determine the mean performance value of those who took the course. Ultimately, the unweighted TOPSIS method was used to evaluate the efficacy of the various training courses in relation to the four criteria. This approach effectively addresses the challenge of using extensive datasets within a system while facilitating the application of a multi-criteria unweighted TOPSIS method. The results of the multi-criteria TOPSIS method indicated that training courses related to the professional fields of administration and management, hostel and tourism and community and sociocultural services have positive impact on employability and improving the working conditions of citizens. However, courses that demonstrate the greatest effectiveness in ranking are the least demanded by citizens. The results will help policymakers evaluate the effectiveness of each training course offered by the regional government.
\end{abstract}

\section{Introduction}\label{sec1}

Lifelong learning is a core concept that describes individual learning during the entire lifecycle from early socialisation and pre-school education to retirement age in terms of employment \citep{Gal2007}. The term lifelong learning is broadly defined and refers to `all learning activity undertaken throughout life, to improve knowledge, skills, and competencies within a personal, civic, social and/or employment-related perspective' \citep{informeCE}. This concept addresses three fundamental objectives of education: i) personal fulfilment and development throughout life (cultural capital); ii) active citizenship and inclusion (social cohesion); and iii) employability and economic growth (human capital). In this study, we focus on the impact of lifelong learning on the third objective, namely, the working life of both employed and unemployed people, which is considered essential either for obtaining a job or for improving the job one already has. In this regard, many national and regional governments in Europe dedicate part of their policy budgets to providing training courses that can offer new skills to their citizens.

The importance of lifelong learning for improving employability and the quality of employment is evident. Numerous studies have confirmed the positive effect of lifelong learning education on improving employability. One example is the study by \cite{Jarlstrom2020}, which asserts that a highly skilled and knowledgeable worker is an asset to any organisation, and skillsets are often associated with promotion, salary increases, and career success. \cite{Babos2015} also concluded that most individuals who participated in lifelong learning recognised it as a positive contributor to their employability. The work of \cite{Sharma} uses micro-credential courses for upskilling and reskilling, demonstrating their positive impact on enhancing employability. To achieve this, interviews were conducted with 65 participants from India, Nigeria, the United Arab Emirates, and the United Kingdom to explore how micro-credentials can be a valuable addition to the higher education ecosystem. Therefore, to provide quality training, it is crucial to have the necessary tools to measure the effect of these educational training courses on employment.

The accelerating pace of scientific and technological advancements and the resulting societal and economic (or labour market) changes at any given time necessitate lifelong learning. This is why policymakers continually face the problem of ensuring that individuals acquire the relevant skills and knowledge to improve employability. Hence, there is a need to design and select education training courses that adapt to the requirements of the labour market. To address this issue, the public employment service of the region of Extremadura (Spain) provided historical data and requested a scientific analysis of the impact that the realisation or not of different continuous training courses has on the working lives of those enrolled. The key contribution of this work is the development of a multi-criteria decision support system that will help policymakers evaluate each training course offered by the regional government and rank them according to improvement in the labour life of students. This is important because only with precise data can the regional government determine where more resources should be invested and which courses are less efficient. The proposed methodology combined the challenges of using large amounts of data in the system and the application of a multi-criteria unweighted method.

One of the most commonly used multi-criteria methods is the TOPSIS method, which is based on distances to ideal and anti-ideal solutions. In TOPSIS, as in other multi-criteria decision (MCDM) methods, the objective is to help decision makers choose the best option among several alternatives based on various criteria that may conflict with one another. These criteria may also have different levels of importance. Therefore, assigning criteria weights is a crucial step in any MCDM method. The weights can significantly affect the final decision even with slight changes. A common method of determining weights is the use of expert opinion. Experts in the field can define the preferences of the criteria using specific methodologies, such as reaching an agreement (Delphi method \cite{Sackman1974}) or comparing criteria in pairs (AHP \cite{Saaty2012}). However, these expert-based methods have a major drawback: they introduce subjective views that can bias the process.

Many researchers in the field of multi-criteria analysis have been interested in finding more objective methods for determining criteria weighting. Examples of these methods include the Entropy method \citep{Shannon1948}, the LINMAP method \citep{Srinivasan1973}, and recent approaches such as the IDOCRIW method \citep{Zavadskas2016},  the Bayesian approach \citep{Vinogradova2018} and the FUCOM method \citep{Pamucar2018}.

A different approach to address the issue of weight assignment in the TOPSIS method was proposed by \cite{Liern2022} and \cite{Benitez2021}. This alternative approach does not require decision makers to specify the exact values of the criteria weights. Instead, it only requires providing reasonable ranges for them, leading to an unweighted TOPSIS (uwTOPSIS) method.

This remainder of this paper is structured as follows. In Section 2, we review related works on the criteria selected to measure the effect of training courses and the selection of the control group to measure that effect, and other studies using uwTOPSIS as multi-criteria decision analysis method in different scientific domains. In Section 3, we describe the database and the courses considered in this study. In Section 4, we present the methodology, define the working life curves, describe how the control groups are defined and explain how the effectiveness of the courses is assessed. In this section, we also briefly describe the uwTOPSIS method used for ranking the courses. The results are presented in Section 5, and we provide the conclusions in Section 6. 

\section{Related works}

Analysing the effect of a certain factor or treatment on a sample is common in many scientific fields. In such analyses, it is essential to define a control group to contrast the results. For example, in pharmacy, a control group helps determine whether a new medication is effective and in agronomy, it is used to study the effects of different fertilisers. In these scientific fields, it is common for researchers to have control over both the sample they are studying and the control group that they use to compare results. In this way, it is possible to select homogeneous samples with similar individuals so that the comparisons make sense and the possible differences can be attributed only to the factor or treatment being analysed. 

However, in social sciences, the situation is often much more complicated. For example, when assessing the effects of participating in a certain activity on individuals, researchers may have limited control over the sample being analysed, which can lead to heterogeneity among participants. In addition, in some studies, the control group has not been previously defined, making the comparison of results with those of individuals who have not participated in the activity a nontrivial process.

In this study, we present a methodology that allows us to assess the impact that the realisation or not of different continuous training courses, offered by the public employment service of the region of Extremadura (Spain), has on the working life of the people enrolled in them. To evaluate the effectiveness of training courses, it is necessary to determine measurable criteria. Many studies have based the evaluation of the criteria on the subjective perceptions of actors such as students, teachers and organisers \citep{Kirkpatrick2006, Farjad2012, Sharma}. In contrast, the proposed methodology is entirely data-driven. This involves constructing a database that includes information from various departments of the regional government, such as education, labour and social security

When assessing the usefulness of a training course, one might be tempted to analyse the quality of working life before and after taking the course and to check whether there has been improvement, according to predefined criteria. However, this can be misleading for two main reasons: first, the quality of working life is likely to improve over time, and second, it is crucial to compare these improvements with those of individuals who have not completed the course. In other words, how much (if any) improvement can be attributed to the training course, or can it all be explained by the natural improvement in the working life of a person? 

To accurately determine whether the obtained results were due to the course, it is necessary to compare the results with those of a control group consisting of individuals who have not completed the course. Several authors have addressed the problem of determining the control group for this type of study in various ways. For example, \cite{Rotar2021} studied the evaluation of employment programmes in the Netherlands, with the control group defined as individuals in the same age range as the study group who did not participate in any employment programme in 2008. This study was limited to a specific period (only 2008 was considered). In contrast, another study \citep{Elena2014} considered the effectiveness of vocational training and defined the control group as individuals who enrolled in the training course but did not attend (no-shows). In other cases \citep{Towler2019}, the analysis focused on a specific type of course in which learning certain skills was evaluated. In these instances, the determination of the control group is simpler because it is sufficient to consider individuals who took similar courses in which that skill was not taught. Similarly, in another study \citep{Sanulita}, the effectiveness of audiovisual learning programmes was analysed using an experimental design with a post-test-only control group. The study involved two classes: the experimental group that received the specific treatment and the control group that did not.

Thus, determining the control group is not a simple task. This presents two major challenges. The first is how to define a control group in a manner that ensures that similar individuals are compared and that the comparison is sensible. This is difficult because, in our case, the students taking the course are not a homogeneous group; they have different backgrounds, ages, and other characteristics. Consequently, the construction of the control group should be individualised, meaning that a specific control group should be defined for each individual taking the course. The second challenge is even more complex. If the people in the control group have not taken the course (by definition), how can we establish a valid ‘before’ and ‘after’ comparison?

To address these challenges, we first define a \textit{`working life curve'} (WLC) that allows us to quantitatively describe the working life of an individual. Using these WLCs, we can measure the similarity between two distinct individuals. In addition, using a methodology based on K-medoids clustering \citep{Park2009}, we define a control group for each individual to facilitate comparison. The performance of each course, in relation to the four predefined criteria, is calculated as the average performance of the individuals who took the course, as assessed using a t-test comparison. Finally, we propose the uwTOPSIS method to rank the effectiveness of different training courses according to four criteria \citep{Benitez2021, Liern2022}.

The reason for selecting the uwTOPSIS multi-criteria method over the existing methods was the challenge of determining precise subjective weights from experts. The proposed multi-criteria method ranks decision alternatives based on the classical TOPSIS approach; however, this method does not require the introduction of fixed prior weights. Instead, it uses lower and upper bounds to express the varying importance of the criteria, thereby allowing for more objective assessment without the bias introduced by subjective weighting. Developed in 2020, this innovative multi-criteria method differs from other known multi-criteria techniques, such as VIKOR, PROMETHEE or MOORA, as it does not assign fixed weights a priori to a criterion; instead, upper and lower bounds are set. 

Recent studies related to the case study have applied the uwTOPSIS method for the reasons explained above. For example, \cite{Blasco} proposed an academic performance indicator for science and engineering students at the Industrial University of Santander (Colombia). Similarly, \cite{Lopez-Garcia} used the uwTOPSIS method to develop a methodology for the early detection of student failure. Other studies in different fields, such as sustainability and tourism, have also demonstrated the efficiency of the uwTOPSIS method in multi-criteria decision-making processes \citep{Blasco-Blasco2021,Perez-Gladish2021,Liern2021,Lopez-Garcia2023}.

\section{Data}\label{sec:data}

The data used in this study are a subset of the data stored in a data warehouse that we built in collaboration with the regional government of Extremadura to support a data-driven strategy aimed at reducing unemployment in the Extremadura region \citep{conejero2021a,conejero2021b}.

All details regarding the construction of this data warehouse, including the identification of the data sources, the description of the data collected, the design of the data warehouse schema and the creation of an automated data collection process, are detailed in Section 3 of \cite{conejero2021a}.

In summary, this data warehouse contains information on 120,927 citizens of the Extremadura region. Specifically, this information for each citizen in the data warehouse includes the following:
\begin{itemize}
    \item Educational from lower and upper secondary, vocational, and official language schools
    \item University degrees (bachelor’s, master’s and doctorate) awarded by the University of Extremadura
    \item Contracts, social security and visits to the Employment Office
    \item Training courses offered by the Employment Office, aimed at helping individuals easily enter and remain in the job market.
\end{itemize}

\subsection{Datasets}

In this study, a subset of the information stored in the aforementioned data warehouse was used. In particular, the information was obtained from the following four different datasets, with their fields described in Table~\ref{tab:datasets}:

\begin{enumerate} 
\item[(a)] \textbf{DS1} This dataset includes citizens who have not participated in any training courses. It contains a row for each regulated study level of each citizen who has not taken a training course. For instance, a citizen with lower secondary education and bachelor’s, and master’s degrees will appear four times in this view.
This table contains 196,120 rows corresponding to studies out of 112,638 different citizens.

\item[(b)] \textbf{DS2}. This dataset includes citizens who have participated in training courses. It contains a row for each regulated study level of each citizen who has taken at least one training course. For example, a citizen with lower secondary education and three training courses will appear five times in this view.
This table contains 21,970 rows corresponding to studies of 8,289 different citizens.

\item[(c)] \textbf{DS3}. This dataset includes personal and aggregated data of all the citizens considered. It contains one row for each of the 120,927 citizens that appear in the two previous views.

\item[(d)] \textbf{DS4}. This dataset includes the contract history for each citizen. It contains one row for each contract held by the 120,927 citizens. 
This table contains 820,985 rows. It should be noted that the number of contracts differs significantly between citizens with and without training courses. Thus, out of the 8,289 citizens with training courses, 8,151 appear at least once with a contract, whereas 138 do not. Conversely, of the total of 112,638 citizens without training courses, 67,253 appear at least once with a contract, whereas 45,385 do not.

\end{enumerate}

\begin{table}[ht]
\caption{Data fields per dataset and their description}\label{tab:datasets}
\centering
\small
\begin{tabular}{c|c|c|c|l|p{7cm}}
  \toprule
DS1 & DS2 & DS3 & DS4 & Name & Description  \\ 
  \midrule
  \checkmark & \checkmark & \checkmark & \checkmark & citizenId       & Unique identifier for each citizen \\
  \checkmark & \checkmark &            & \checkmark & endDate         & Date of degree (DS1 and DS2) or contract (DS4)  \\
  \checkmark & \checkmark &            &            & studyType       & Compulsory, vocational, university or training course \\
  \checkmark & \checkmark &            &            & degree          & Concrete name of the degree obtained  \\
             &            & \checkmark &            & gender          & Citizen's gender  \\
             &            & \checkmark &            & birthDate       & Citizen's date of birth  \\
             &            & \checkmark &            & age             & Citizen's age  \\
             &            & \checkmark &            & numberOfStudies & Total number of studies of the citizen  \\
             &            & \checkmark &            & daysOfWork      & Days worked by the citizen  \\
             &            &            & \checkmark & typeCode        & Code of the type of contract  \\
             &            &            & \checkmark & description     & Description of the contract type  \\
             &            &            & \checkmark & startDate       & Start date of the contract  \\
             &            &            & \checkmark & typology        & Contract type (temporary or permanent)  \\
             &            &            & \checkmark & cnoCode         & Occupation National Code (CNO in Spanish)  \\
             &            &            & \checkmark & cnoDesc         & Occupation description according to the CNO  \\
             &            &            & \checkmark & cnaeCode        & Economic Activity National Code (CNAE in Spanish)  \\
             &            &            & \checkmark & cnaeDesc        & Economic Activity Description according to its CNAE  \\
             &            &            & \checkmark & economicSection & Economic section where the contract is set  \\
             &            &            & \checkmark & sector          & The sector where the contract is set  \\
             &            &            & \checkmark & localityCode    & Code of the locality where the contract takes place  \\
             &            &            & \checkmark & pfCode          & Code of the professional family for which the contract is classified  \\
   \bottomrule
\end{tabular}
\end{table}
 
\subsection{Courses}

For this study, 113 training courses offered by Extremadura’s employment service (SEXPE) in Spain were considered. Because we set a study horizon of 1 year and were interested in assessing the impact of training courses, we filtered the data to keep only those citizens who completed a training course more than one year before data collection, making a total of 6748 citizens considered. The number of students per course ranges from 10 to several hundred students depending on the course. The basic descriptive statistics of the number of students in each course are presented in Table \ref{tab:coursestats}. 
In addition, all courses are assigned to a professional family that each job contract is classified into. Assignments were completed according to the proximity of the course description to the corresponding professional family. The list of professional families and their corresponding codes are presented in Table \ref{tab:famprofcodes}.

\begin{table}[ht]
\caption{Number of students enrolled in the different training courses under study.}\label{tab:coursestats}
\centering
\small
\begin{tabular}{rrrrrr}
  \toprule
Min. & 1st Qu. & Median & Mean & 3rd Qu. & Max. \\ 
  \midrule
4.00 & 12.00 & 19.00 & 61.51 & 41.00 & 961.00 \\ 
   \bottomrule
\end{tabular}
\end{table}

\begin{table}[ht]
\caption{Professional family codes. Every course has been assigned to a professional family.}\label{tab:famprofcodes}
\centering
\small

\begin{tabular}{llll}
  \toprule
Code & Description & Code & Description \\ 
  \midrule
ADM & Administration and Management & HEA & Health \\ 
ALA & Agricultural and Livestock Activities & HOT & Hostel and Tourism \\ 
ART & Arts and arts craft & IMA & Installations and Maintenance \\ 
BCW & Building and Civil Works & IMS & Image and Sound \\ 
CHE & Chemistry & ITC & IT and Communications \\ 
COM & Commerce and Marketing & MEM & Mechanical Manufacturing \\ 
CSS & Community Sociocultural Services & PIM & Personal Image \\ 
ELE & Electricity and Electronics & PSA & Physical and Sports Activities \\ 
ENW & Energy and Water & TCL & Textile, Clothing and Leather \\ 
FOI & Food Industries & VTM & Vehicle Transport and Maintenance \\ 
GRA & Graphic Arts & WFC & Wood Furniture and Cork \\ 
   \bottomrule
\end{tabular}
\end{table}

\section{Methodology}\label{sec:method}
\begin{figure}[htpb]
\centering
\includegraphics[width = .8\linewidth]{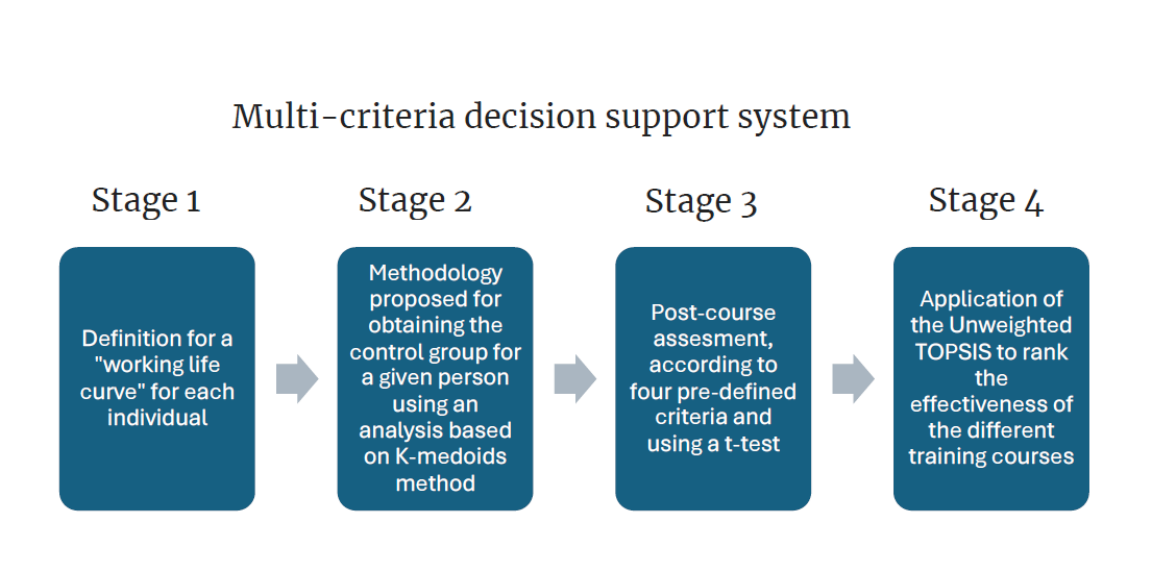}
\caption{Proposed methodology for developing a multi-criteria decision support system}\label{fig:Meth}
\end{figure}

To assess the impact of a training course on a citizen’s employment track, we propose a measure to compare the working track of different citizens: the WLC. 
Based on this measure, we then define a method to compute a control group for each citizen by clustering the citizens according to their WLCs. Finally, we discuss how to assess the impact of a training course according to multiple criteria. The methodology is summarized in four stages, as shown in Figure \ref{fig:Meth}.

\subsection{Working life curves}
We define the WLC of person $i$ in the database as a function $WLC_i: [0,T_{\max}]\longrightarrow [0,1]$ that assigns to each day $t$ of the working life of person $i$ the quotient between the number of days that a person has been employed and the number of days spanned since the beginning of their working life. That is, if $N(t)$ is the number of days under employment until day $t$, then we have the following:
\begin{equation}\label{eq:wlc}
WLC_i(t) = \dfrac{N(t)}{t},\quad t = 1, 2, \ldots    
\end{equation}

\begin{figure}[htpb]
\centering
\includegraphics[width = .8\linewidth]{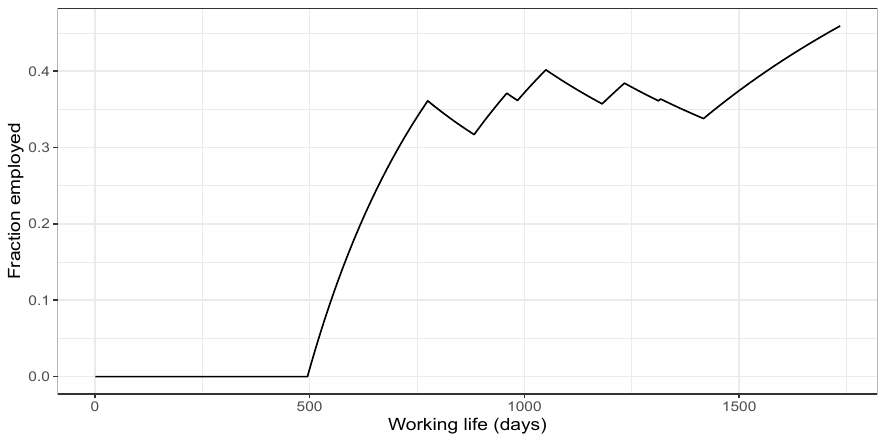}
\caption{Example of a working life curve (WLC) for a person. An increasing WLC indicates that the person is employed.}\label{fig:WLC1}
\end{figure}

An example of a typical WLC is shown in Figure \ref{fig:WLC1}.
 
When constructing the WLC, two points are convenient to emphasise:

\begin{itemize}
\item Determination of the date of origin of working life. Normally, the beginning of working life is defined as the earliest between the start date of the first labour contract and the end date of the last regular study taken (not including the possible SEXPE training courses taken). 
\item In many cases, the end date of a contract is not recorded and appears in the database as \texttt{NULL}. More than 25\% of the contracts do not have an appropriately assigned end date. Therefore, a mechanism to assign an end date to the contract must be established.

In this case, we proceed as follows:

\begin{enumerate}
\item We consider as adjusted end date the midpoint of the quarter following the start of the contract. That is, if \texttt{month} denotes the month in which the contract started, then we have the following:
\begin{itemize}
\item If \texttt{month} $\in\{\text{January, February, March}\}$, then \texttt{date\_adjusted}$=$ 15th of May.
\item If \texttt{month} $\in\{\text{April, May, June}\}$, then \texttt{date\_adjusted}$=$ 15th of  August.
\item If \texttt{month} $\in\{\text{July, August, September}\}$, then \texttt{date\_adjusted}$=$ 15th of  November.
\item If \texttt{month} $\in\{\text{October, November, December}\}$, then \texttt{date\_adjusted}$=$ 14th of February of the following year.
\end{itemize}
\item The beginning date of the next contract (if any) is also considered.
\item The final date is considered the earliest between the two above dates: the earliest date between \texttt{date\_adjusted} and the date at the beginning of the next contract. 
\end{enumerate}

It may happen that reality does not conform to this methodology and that there are reasons why no contract end date was communicated. However, in our opinion, this methodology is a systematic and sensible way to assign end dates.
\end{itemize}

\subsection{Definition of the control group}
The next step is to compare the WLC of person $p_i$ who has completed a particular training course with that of other people who have not. In particular, we want to determine whether from the completion of a course, at instance $t_i$ (measured in days), the working life of an individual improves in horizon $h$, that is, in the interval $[t_i, t_i+h]$, when compared with a group of people who did not take that course and, up to moment $t_i$, had a labour behaviour similar to that of the individual in question. 

Assume that from a course $C_j$ under study, we take an individual $p_i$ who has taken the course at instance $t_i$ (measured in days since the beginning of his or her working life). Note that, in addition to $t_i$, for each person in the database, we know the sex, age and educational level at each instance. The control group for $p_i$ (denoted $CG_i$) is obtained in three stages. 

First, we filter the database by considering only individuals of the same sex as $p_i$, age within 5 years of the age of $p_i$, the same educational level at time $t_i$ as person $p_i$, and working life at least as long as that of $p_i$ plus horizon $h$ (in days). This defines the `\textit{initial control group}' for $p_i$ (denoted$CG_{i0}$). 

However, given the massive size of the database, the initial control group $CG_{i0}$ is expected to be large. Therefore, a second filter is required. The aim is to select from $CG_{i0}$ individuals whose WLCs are most similar to the WLC of the person under investigation. The next problem is determining the number of `nearest neighbors' to $p_i$ because this depends heavily on the course and the individual in question. 

To address these issues, we propose a methodology based on K-medoids clustering \citep{Park2009}. The general idea is to consider, given person $p_i$, the WLCs corresponding to $CG_{i0}\cup \{p_i\}$, all of them in the interval $[0, t_i]$. Then, the control group of $p_i$ (denoted $GC_i$) is the optimal cluster containing $p_i$ for $K$-medoids clustering. The optimality is determined using the GAP statistic \citep{Tib2001}, and the PAM algorithm \citep{Sch2019} is used for the clustering procedure. 

Nevertheless, we must note that the size of the initial control group is in the thousands, or even in the tens of thousands. Thus, obtaining the GAP statistic for each person is computationally extremely expensive. To overcome this challenge, pilot tests were conducted, and the optimal number of clusters did not vary greatly from one individual to another. Therefore, instead of determining the optimal number of clusters for each participant who took a specific training course, we computed such an optimal number for a randomly selected sample and took the mode as the optimal number of clusters for the given training course. This number was used to determine the control groups of all participants in the training course.  

\begin{figure}[htpb]
\centering
\includegraphics[width = .8\linewidth]{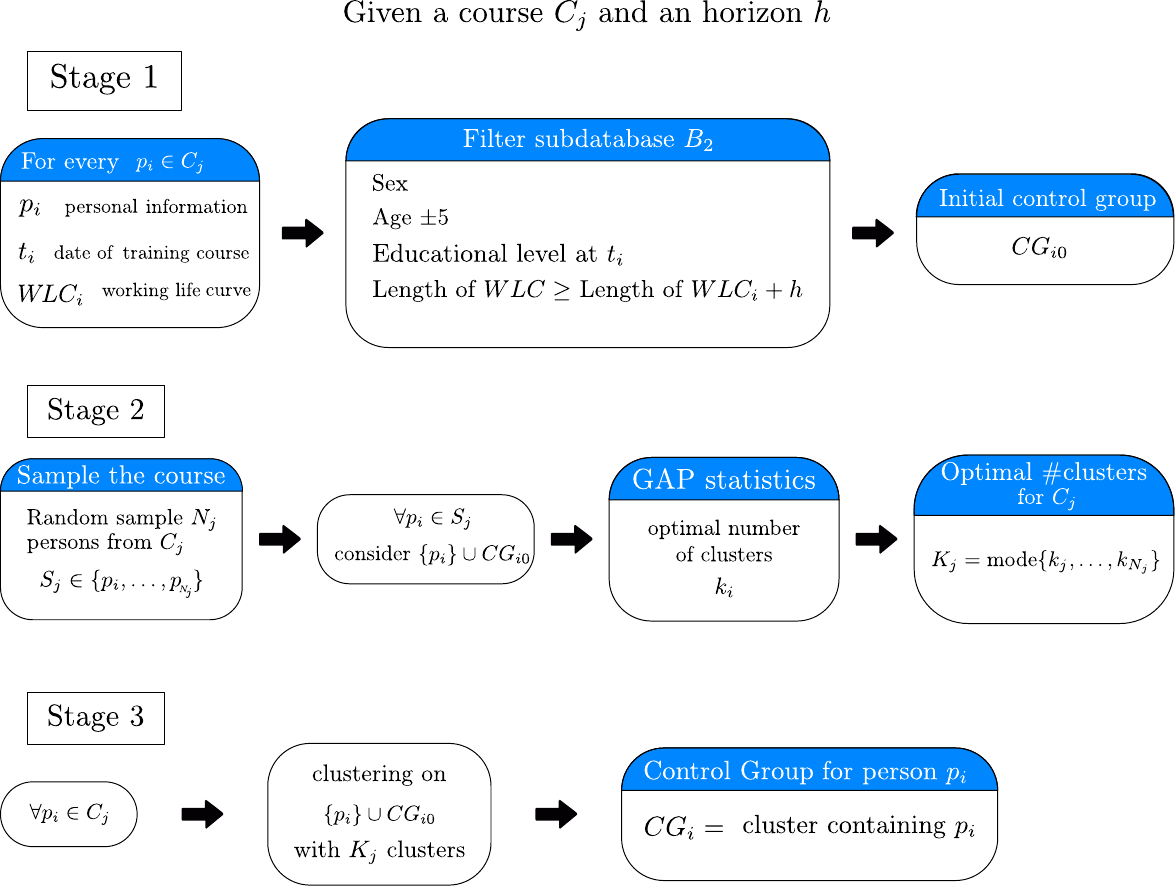}
\caption{Methodology of obtaining a control group for a given person taking course $C_j$.}\label{fig:controlgroup}
\end{figure}

Figure \ref{fig:controlgroup} summarises the three stages involved in determining the control group for a specific person $p_i$ taking course $C_j$. Algorithm 1 describes the steps to be followed to obtain the control group of all individuals who have completed a given course.

\begin{algorithm}\label{alg:CG}
\caption{Determination of control groups of all persons who have taken the course $C_j$}
\begin{algorithmic}[1]
\Procedure {InitialCG}{$C_j$, $p_i$, $h$, $B_2$}
\State Compute $WLC_i$ following \eqref{eq:wlc}.
\State Find $t_i$ (date of the course in days).
\State $a_i,\,s_i\,e_i\leftarrow $ age, sex, ed. level at $t_i$ of $p_i$.
\State $CG_{i_0}\leftarrow$ Filter $B_2$: age $=a_i$, sex $=s_i$, ed. level $=e_i$, length of $WLC\geq t_i + h$.
\State \textbf{return:} $CG_{i_0}$ 
\EndProcedure
\Statex

\Procedure{OptimalClusterNumber}{$C_j$, $N_j$, $h$, $B_2$}
\State $S_j\leftarrow$ Random Sample of $C_j$ with size $N_j$.
\ForAll {$p_l \in S_j$}
\State $CG_{l_0} \leftarrow \Call{InitialCG}{C_j, p_l, h, B_2}$
\State $k_l\leftarrow \textsc{GAP}(\{p_l\}\cup CG_{l_0})$
\EndFor
\State \textbf{return:} $\text{mode}\{k_1,\ldots,k_{N_j}\}$
\EndProcedure
\Statex
\Procedure{ControlGroup}{$C_j$, $N_j$, $h$, $B_2$, $h$}
\State $k_j\leftarrow \Call{OptimalClusterNumber}{C_j,N_j,h,B_2}$
\State $CG(C_j)\leftarrow\emptyset$
\ForAll {$p_i \in C_j$}
\State $CG_{i_0}\leftarrow\Call{InitialCG}{C_j, p_i, h, B_2}$
\State $\text{Kmed}\leftarrow \textsc{KMedoids}(\{p_i\}\cup CG_{i_0})$
\State $CG_i\leftarrow $ cluster in Kmed containing $p_i$
\State $CG\leftarrow CG\cup CG_i$
\EndFor
\State \textbf{return:} $CG$
\EndProcedure
\end{algorithmic}
\end{algorithm}

\subsection{Post-course assessment}

Once we have devised a methodology for comparing individuals, we are now in a position to measure the extent to which the training course has had a positive effect on a particular person in some respect, or not. 

Now, let us consider a particular course $C_j$, and a particular person $p_i\in C_j$ with their control group $CG_i$. Assume that $p_i$ has completed the training course at time $t_i$ (measured in days since the beginning of the WLC). We now want to determine whether the labour conditions of $p_i$ improved after the course, within some horizon $h$ (i.e. in the interval $[t_i, t_i + h]$), compared with the improvement in the persons in the control group $CG_i$ during the same period. 

The improvement above will be measured according to four criteria. Namely, 
\begin{itemize}
\item \textbf{C1:} Total number of days employed 
\item \textbf{C2:} Total number of days under permanent contract
\item \textbf{C3:} Number of days in a position related to the course taken
\item \textbf{C4:} Average number of days between contracts 
\end{itemize}

Criteria C1, C2 and C4 analyse the employability of a training course in general, regardless of its professional family, whereas criterion C3 specifically analyses the employability within the professional family of the course under consideration. With this criterion, we want to analyse the effectiveness of the courses as a facilitator of improvements in working conditions in their own professional family. We include both general and specific conditions because we note that, in many cases, it may happen that courses, although related to one professional family, provide students with resources and transversal skills that are useful in other sectors.

To determine the degree to which a course has been effective for an individual, for a given criterion, we calculate the probability with which we can be sure that the individual has performed better on that criterion than his or her control group. To this end, we perform a t-test to determine whether the value obtained from the individual on a criterion was statistically superior (i.e. higher for criteria C1, C2 and C3 and lower for criterion C4) than those of the control group. 

We use the P-value of the corresponding t-test as the performance measure. This implies that the criteria are of cost type—that is, the higher the value, the worse the performance.
Finally, the course performance is calculated according to the criteria as the average performance of the participants.

\subsection{Unweighted TOPSIS}
\label{sec:topsis}

In this section, we briefly describe both the classical TOPSIS method and the uwTOPSIS formulation proposed by \cite{Liern2022} and the implementation described in \cite{Benitez2021}.

Consider a multi-criteria problem with $n$ alternatives $A_i$, $i = 1,\ldots, n$, and $m$ criteria $C_j$, $j = 1,\ldots,m$. 
Each criterion may be beneficial (i.e. \textit{`the more the better'}) or a cost (i.e.  \textit{`the less the better'}). Let $J_+$ and $J_-$ denote the sets of indices $j$ corresponding to benefit and cost criteria, respectively. 

The classical TOPSIS algorithm proceeds as follows \citep{Hwang1981, Haang2011}: 
\begin{itemize}
\item[{\sc S} ] \hspace{-0.35cm}{\sc tep 1:} Determine decision matrix $X=[x_{ij}]_{n\times m}$ where element  $x_{ij}$ is the performance rating of alternative $A_i$ at criterion $C_j$.

\item[{\sc S} ] \hspace{-0.35cm}{\sc tep 2:} Construct the normalised matrix  $\hat{X}$ as follows:
\begin{equation*}
  r_{ij} \frac{x_{ij}}{\sqrt{\sum_{k = 1}^n x_{kj}^2}}\in [0,1],\quad 1\leq i \leq  n, \, \,1\leq j\leq m. 
\end{equation*}

\item[{\sc S} ] \hspace{-0.35cm}{\sc tep 3:} Determine the positive ideal solution,  $A^+=(A^+_1, A^+_2, \ldots,A^+_m)$, and the negative ideal (or anti-ideal) solution, $A^-=(A^-_1, A^-_2, \ldots,A^-_m)$,  as follows:
\begin{equation*}
A^+_j = \begin{cases}
\displaystyle\mathop{\mbox{max}}_{1\leq i \leq n}
r_{ij}&\text{if } j \in J_+\\
\displaystyle\mathop{\mbox{min}}_{1\leq i \leq n}
r_{ij}&\text{if } j \in J_-
\end{cases}
\qquad 
A^-_j = \begin{cases}
\displaystyle\mathop{\mbox{min}}_{1\leq i \leq n}
r_{ij}&\text{if } j \in J_+\\
\displaystyle\mathop{\mbox{max}}_{1\leq i \leq n}
r_{ij}&\text{if } j \in J_-
\end{cases}
\end{equation*}

\item[{\sc S} ] \hspace{-0.35cm}{\sc tep 4:} Given a weight vector $\omega\in \Omega$, where
\begin{equation*}
\Omega=\left\{(\omega_1, \omega_2, \ldots, \omega_m)\in \mathbb R^m, \quad 0\leq \omega_j\leq 1, \quad \sum_{j = 1}^m\omega_j=1\right\},
\end{equation*} 
we calculate the weighted normalised matrix, $[w_j{r}_{ij}].$

\item[{\sc S} ] \hspace{-0.35cm}{\sc tep 5:}  Determine the weighted distances  between each alternative $A_i$ and the ideal and anti-ideal solutions as follows:
\begin{equation*}
d_i^+({\omega}) = \sqrt{\sum_{j = 1}^m (\omega_j{r}_{ij} - \omega_jA^+_j)^2}, \quad
d_i^-({\omega})=  \sqrt{\sum_{j = 1}^m (\omega_j{r}_{ij} - \omega_jA^-_j)^2}.
\end{equation*}
\item[{\sc S} ] \hspace{-0.35cm}{\sc tep 6:} Compute the score  for each alternative as follows:
\begin{equation}\label{eq:score}
R_i ({\omega}) = \frac{d_i^-({\omega})}{d_i^-({\omega})+d_i^+({\omega})},\quad i = 1,\ldots,n.
\end{equation}

\item[{\sc S} ] \hspace{-0.35cm}{\sc tep 7:} Rank alternatives according to scores $R_i({\omega}) $. 
\end{itemize}

\begin{remark}
In this study, following the original proposal in \cite{Hwang1981}, we use vector normalisation in Step 2. However, other normalisation procedures have also been successfully applied \citep{Jamal, Cables}. Likewise, the Euclidean distance in Step 5 can be replaced by many other distances \citep{Jamal}.
\end{remark}
\medskip

TOPSIS is very easily implemented and applied; however, weight determination (Step 4) is a significant concern because a small change in any of them can lead to different final rankings. Inspired by \cite{Liern2022}, the proposed method attempts to avoid subjective weight assignment by treating weights as decision variables. For this purpose, we consider vector $\omega$ (Step 4) as a vector of decision variables. Therefore, for each alternative, expression \eqref{eq:score} defines a function as follows:

\begin{equation}
\label{funcion}
R_i:  \Omega \rightarrow [0, 1], \quad  i=1, \ldots, n.
\end{equation}
Given the extreme values of the function $R_i$ in \eqref{funcion} for each alternative,  we can define a ranking. Thus, we must to compute the following:
\begin{align}
R^-_i &= {\min} \left\{ R_i(\omega):\quad  
 \sum_{j = 1}^m\omega_j=1,  \quad 
 l_{j}\leq \omega_j\leq u_{j},\quad 1\leq j \leq m
\right\}\label{eq:prog_min2}\\
 R^+_i &= {\max} \left\{ R_i(\omega):\quad  
 \sum_{j = 1}^m\omega_j=1,  \quad 
 l_{j}\leq \omega_j\leq u_{j},\quad 1\leq j \leq m
 \right\}.\label{eq:prog_max2}
\end{align}
Note that parameters $l_j$ and $u_j$, $j =1\ldots, m,$ are lower and upper bounds for weights, respectively. All $l_j$ values should be positive; otherwise, some criteria will not be considered. However, the upper bounds should not be very large because assigning a large weight to one criterion has similar consequences to ignoring the other criterion.

From \eqref{eq:prog_min2} and \eqref{eq:prog_max2}, we define the following interval for each alternative:
\begin{equation}
\label{intervalo}
 \bar{R}_i=[R^-_i, R^+_i],\quad  i=1, \ldots, n.
\end{equation}
Finally, to rank the different alternatives, Step 7 should be extended to ordering intervals rather than real numbers.

A wide range of methods for ordering intervals $A=[a_1, a_2]$ and $B=[b_1, b_2]$ is available in the literature \citep{Gil, Ramik}. However, in this study, we follow the proposal in \cite{Canos} as follows:
\begin{equation}
\label{orden}
A\succ B \Leftrightarrow \begin{cases}
k_1a_1+k_2a_2>k_1b_1+k_2b_2, &k_1a_1+k_2a_2\neq k_1b_1+k_2b_2\\
a_1>b_1,&k_1a_1+k_2a_2= k_1b_1+k_2b_2,
\end{cases}
\end{equation}
where $k_1$ and $k_2$ are two pre-established positive constants such that $k_1 + k_2 = 1$.

\begin{definition}
Given the alternatives $\{A_i\}_{i=1}^n$ and the set of interval-valued  proximities to the ideal solution  $\{\bar{R}_i\}_{i=1}^n$  given by  (\ref{intervalo}), we conclude that alternative $A_i$ is preferable to alternative $A_k$ whenever $\bar{R}_i\succ  \bar{R}_k$. In addition, we define the {\it uwTOPSIS indicator} $R^{uw}$, of alternative  $A_i$ as follows: 
\begin{equation}
    \label{uwT}
   R^{uw}_i=k_1 R^-_i+k_2 R^+_i, \quad 1\leq i \leq n.
\end{equation}
\end{definition}

The $k_1$ and $k_2$ (or $1-k_1$) values can be considered a measure of the decision maker's propensity to consider a more pessimistic (or conservative) or optimistic scenario. The choice of one or another value of $k_1$ depends on the problem in question and, above all, on the existence of external constraints that may incline the decision-maker towards one of the extremes of the interval obtained in \eqref{intervalo}. In our case, in the absence of such constraints, we rank the intervals by taking $k_1=k_2=0.5$ in (\ref{orden}). Specifically, we use as the uwTOPSIS indicator the midpoints of the intervals $[R^-_i, R^+_i]$ (see equation \eqref{uwT}).

The main drawback associated with this method is the computational difficulty of solving $2n$ nonlinear optimisation problems. To overcome this problem, we used R statistical software \citep{Rcore} together with the package \texttt{nloptr} \citep{nloptr} which is an R port of the open-source nonlinear library developed by S.G. Johnson \citep{nlopt}. The minimising algorithm used to solve problems \eqref{eq:prog_min2} and \eqref{eq:prog_max2} was the Constrained Optimisation by Linear Approximations algorithm developed in \cite{Powell1994}.

\section{Results}\label{sec:results}

\subsection{Control groups}
\begin{figure}
    \centering
    \includegraphics[width = .85\linewidth]{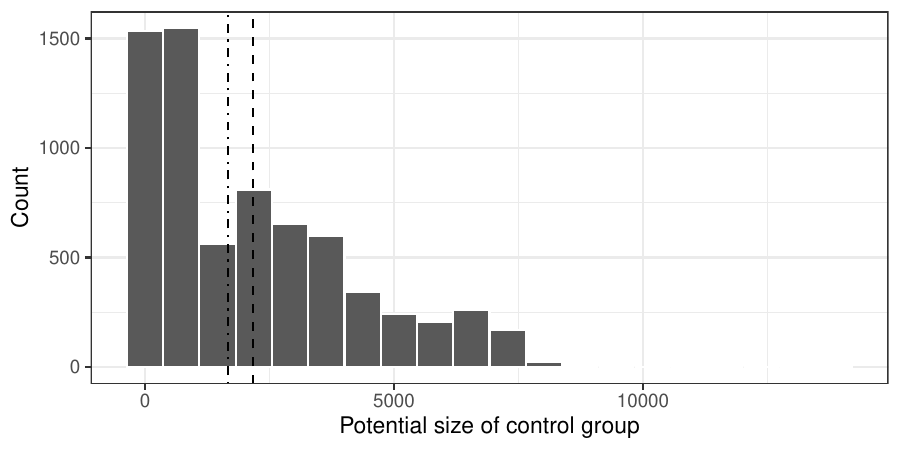}
    \caption{Distribution of the potential sizes of the control groups. The mean (dashed line) and median (dotted-dashed line) are represented by vertical lines.}
    \label{fig:cgsize}
\end{figure}

After the first filtering (sex, age, educational level, and length of working life), the potential size of the control groups had a very skewed distribution to the right, with a mean of 2170 persons and a median of 1662 persons (Figure \ref{fig:cgsize}). 

In all these potential control groups, the Euclidean distance between the individual's labour curve and the labour curves of the persons included in the corresponding potential group was calculated. The potential group was then ordered in increasing order. 

\begin{figure}[ht]
    \centering
    \includegraphics[width = \linewidth]{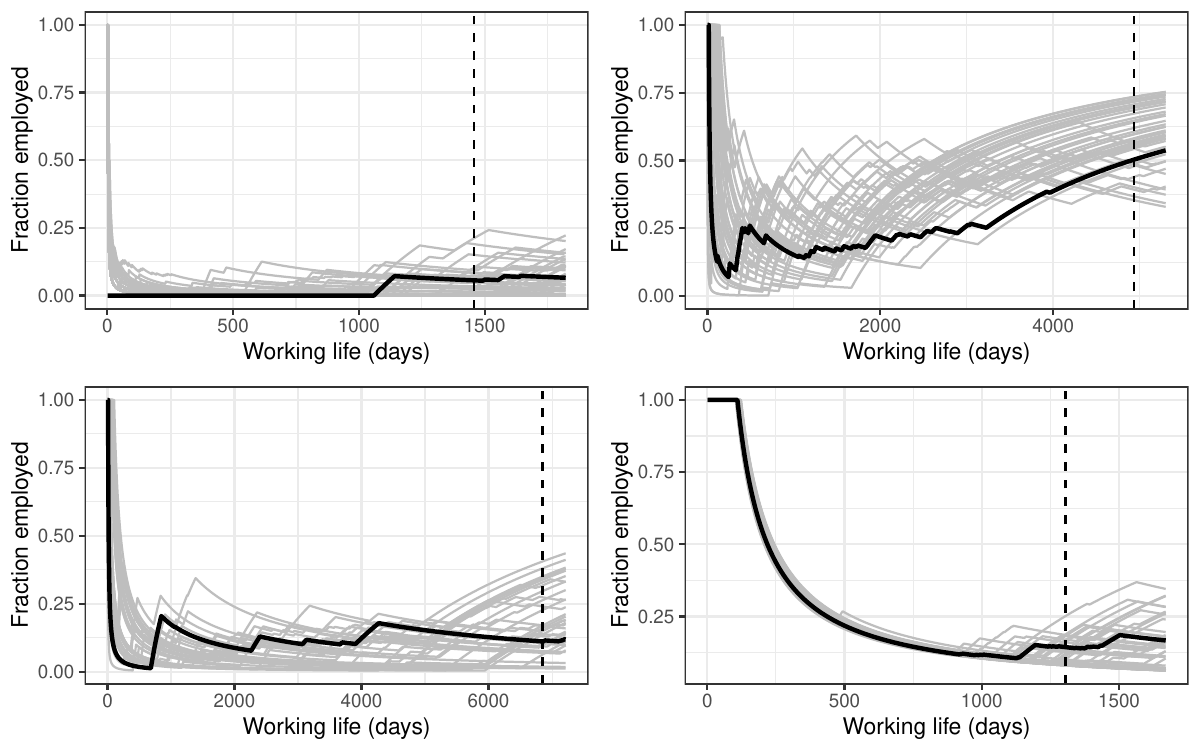}
    \caption{Examples of working life curves of four individuals (black thick lines) and their final control groups (grey lines). The vertical dashed line represents the moment the participants took the course. For clarity, only up to 50 nearest neighbours were plotted.}
    \label{fig:GruposControl2}
\end{figure}

We then proceeded to find the final control groups by k-means clustering and determined the optimal number of clusters using the GAP statistic. Because of the high computational complexity of calculating the GAP statistic, we decided to simplify the calculation by first reducing the potential size of the control group to 500 individuals (i.e. the 500 individuals closest to the person under study). Since it was observed that in almost all cases the optimal number of clusters was between 4 and 5, we decided to perform a random sampling of 200 individuals and found that 5 was the most repeated optimal number of clusters; thus, that value was used for all clusters. Figure \ref{fig:GruposControl2} presents four examples of the outcome of this methodology for defining the control groups. 

\begin{figure}[ht]
    \centering
    \includegraphics[width = .85\linewidth]{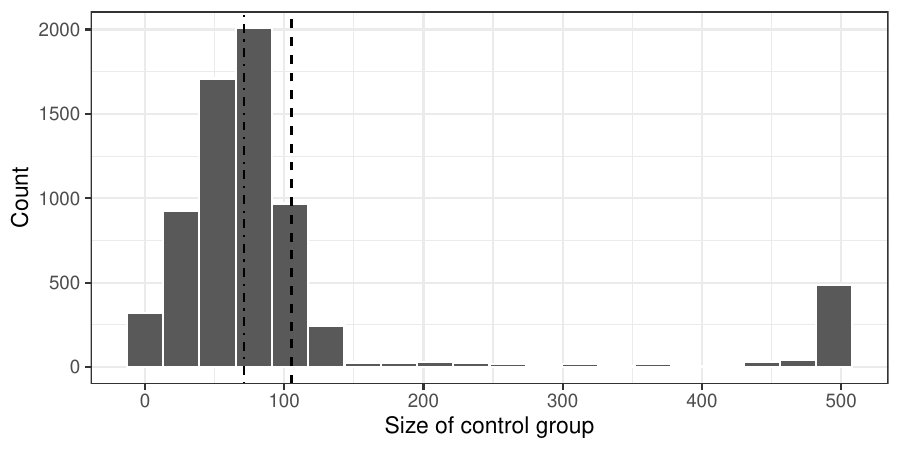}
    \caption{Distribution of the sizes of the final control groups. The mean (dashed line) and median (dotted-dashed line) are represented by vertical lines.}
    \label{fig:controlgroup2}
\end{figure}

After performing clustering and defining the control group of each individual as the cluster to which that individual belongs, Figure \ref{fig:controlgroup2} shows that the sizes of the control groups followed a similar distribution to that of the previous group, with a mean of 105.32 individuals and a median of 71 individuals. In this distribution a high value (specifically 496 individuals) was observed. This is because by limiting the number of individuals to 500 and forcing clusterings of five groups, in some cases in which the potential group was very large (several thousand individuals), the 500 closest people were already very similar to the person under study, so the optimal grouping was to make a group with 496 individuals and then another four groups with a single individual (Figure \ref{fig:controlgroup2}).

\subsection{Criteria assessment}
\begin{figure}[thp]
    \centering
    \includegraphics[width=\linewidth]{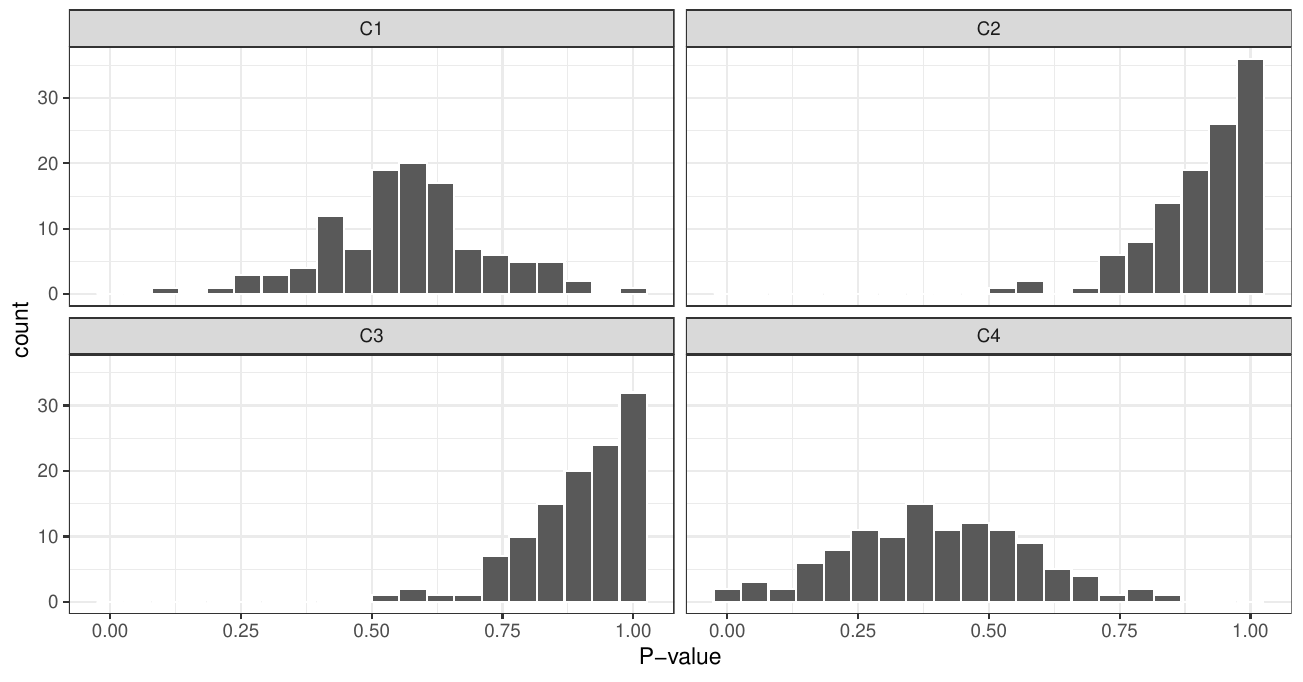}
    \caption{Distribution of P-values for the four criteria considered.}
    \label{fig:Pvalues_distribution}
\end{figure}

The general distribution of course performance for different criteria is shown in Figure \ref{fig:Pvalues_distribution}. It is worth noting that individual scores were, in general terms, quite extreme. Notably, they were either close to 1 or close to 0. Hence, the performance of a course at a certain criterion was highly and positively correlated to the percentage of students of that course that would have not passed a t-test at a standard significance level (e.g. 5\% or 10\%).

As shown in Figure \ref{fig:Pvalues_distribution}, criteria C2 and C3 were rarely met by respondents. The C2 criterion (`total number of days with a permanent contract') is very restrictive in Spain and even more so in regions such as Extremadura, where employment is characterised by high levels of temporary employment. Therefore, the results obtained for different courses under criterion C2 were as expected.

In contrast, the poor performance results for criterion C3 (`number of days in a position related to the course taken'), provided us some useful information. The fact that the results for this criterion were also very poor indicates that, in general, SEXPE training courses do not provide specific training relevant to the Extremaduran labour market.

\subsection{Course ranking}
Once the different alternatives were evaluated for each of the four criteria and the performance matrix was obtained, we ranked them using the uwTOPSIS method described above. In this case, we used $l_j = 0.1$ and $u_j = 0.6$ for $j = 1,\ldots,  4$ as the lower and upper bounds for different criteria, respectively. The rationale behind this selection of limits is twofold. Firstly, the wide range of values reduces the influence of subjective decision-making. Secondly, the ranges must be sufficiently narrow to exclude situations in which some weights approach 0 or 1. In such cases, significant issues may arise during the decision-making process. In the former case, a criterion can be eliminated; in the latter case, the problem is no longer multi-criteria.

Table \ref{tab:top6} presents the ranking results obtained using the uwTOPSIS method. Column `uwTOPSIS' shows the final TOPSIS score, and columns `Min' and `Max' contain the minimum and maximum values, respectively, of the TOPSIS score obtained in the optimisation procedure described by equations \eqref{eq:prog_min2} and \eqref{eq:prog_max2}.

\begin{table}[ht]
\caption{Top 6 courses}\label{tab:top6}
\centering
\small
\begin{tabular}{lrrrc}
  \toprule
 Course & Min & Max & uwTOPSIS & Postition \\ 
  \midrule
 Corporate financing & 0.63 & 0.77 & 0.70 &   1 \\ 
 Creation and manag. package tours \& events & 0.64 & 0.75 & 0.69 &   2 \\ 
 Reception at lodging facilities & 0.62 & 0.73 & 0.67 &   3 \\ 
 Assembly \& storage of refrig. systems & 0.60 & 0.74 & 0.67 &   4\\ 
 Restaurant services & 0.58 & 0.75 & 0.66 &   5 \\ 
 Administr. \& Financ. manag. internat. trade & 0.61 & 0.71 & 0.66 &   6\\ 
 \vdots&&&&\vdots\\
   \bottomrule
\end{tabular}
\end{table}

The aggregated ranking analysis revealed a significant drawback. An analysis of the number of students who took courses reveals that only approximately 6\% of the students took courses in the first quartile, whereas more than 60\% of all students took courses in the third and fourth quartiles (Table \ref{tab:nstudentsquartile}). 

\begin{table}[ht]
\caption{Total number of students in courses in each quartile.}\label{tab:nstudentsquartile}
\centering 
\small
\begin{tabular}{rrrrr}
  \toprule
Q1 & Q2 & Q3 & Q4 & Total \\ 
  \midrule
 404 & 2228 & 2413 & 1703 & 6748 \\ 
 6\% & 33 \% & 36\% & 25\% & 100\% \\ 
   \bottomrule
\end{tabular}
\end{table}

This might be of relevance for policymakers, as it may be an indicator of where public resources should be increased to optimise the effectiveness of training courses. It is possible that some of the most demanded courses are not performing as effectively as expected. With better targeting or more publicity for other courses, citizens would have better information when choosing how to pursue their training to improve their work quality.

\begin{figure}[ht]
    \centering
    \includegraphics[width =.85\linewidth]{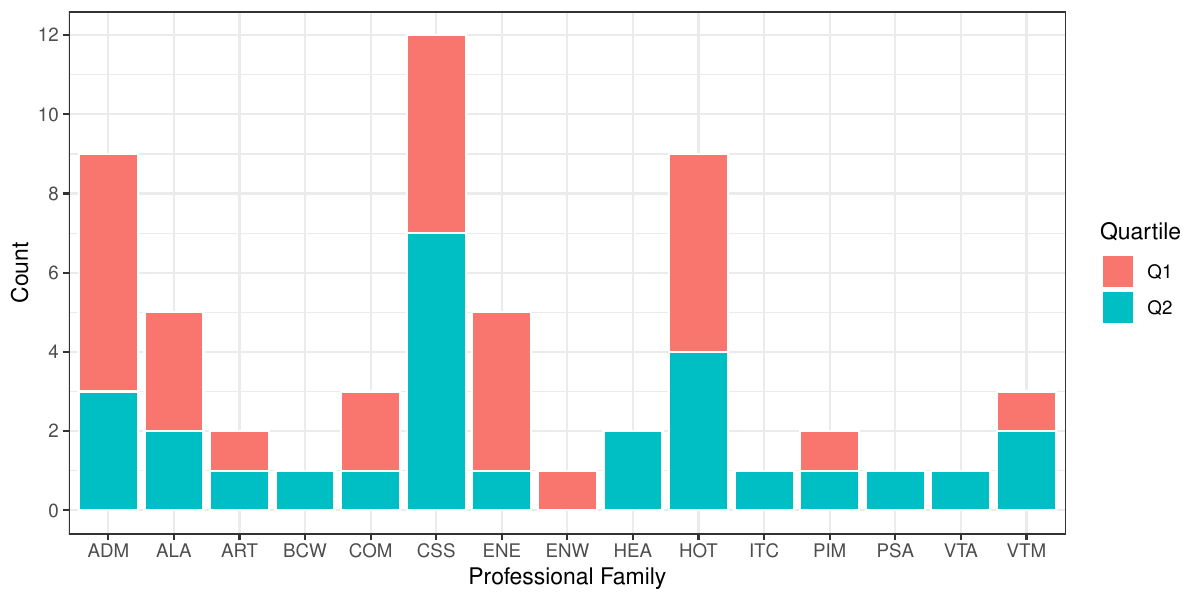}
    \caption{Number of courses in the first two quartiles grouped by professional family.}
    \label{fig:Quartiles12}
\end{figure}

Furthermore, because of the great diversity of courses, they can be conveniently grouped by professional families to determine which types of professional activities perform better in the labour market. Therefore, Figure \ref{fig:Quartiles12} illustrates courses grouped by professional families, appearing in the first two quartiles of the ranking.

The figure highlights three professional families that stood out from the rest: community and sociocultural services (CSS), administration and management (ADM), and hotel and tourism (HOT). The first category is a good indicator of the ageing of the Extremaduran society (and the Spanish society in general, mostly in the inner regions of Spain) because many of the courses within this professional family are related to assistance for the elderly and dependent people.  

In contrast, ADM is generally related to business management; thus, it is not surprising that it exhibited good performance according to the defined criteria. Meanwhile, HOT is linked to hospitality and tourism, which have recently gained a great boost in the region because of the rise in inland tourism and the proximity of Madrid, a major source of tourists.

\subsection{ Sensitivity analysis}

One of the main advantages of using uwTOPSIS over other weighted methods is that, in the absence of expert information, the decision maker only has to provide some rough lower and upper limits for the weights of each criterion, thus decreasing subjectivity during the decision-making process. However, this does not mean that the results are free from subjectivity because they depend on the upper and lower limits defined above and on the value of parameter $k_1$, which defines the inclination of the decision maker towards one extreme or the other of the interval obtained by uwTOPSIS in equation \eqref{intervalo}, and whose usual value is $k_1 = 0.5$. It is thus important to analyse the effect that changing the definition of these parameters has on both the uwTOPSIS score and final ranking.  

To this end, we conducted a simple analysis starting from the values of the parameters used in the previous section (i.e. $l_j = 0.1$ and $u_j = 0.6$, for $j = 1, \ldots, 4$, and $k_1 = 0.5$). We varied these values to a certain extent and measured the difference between the obtained and original solutions. It is essential to carefully consider and maintain objectivity when determining the degree of uncertainty to be assigned to the input factors (in our case, the criteria weights or the parameter $k_1$) to assess the variance of the output (i.e. score or the position in the ranking). If the degree of variation attributed to input factors is excessive, the model exhibits an unacceptable degree of variability, rendering it ineffective \citep{Leamer1983}. In accordance with the methodology proposed by other authors, such as \cite{Annoni2010}, the parameters were varied within the following range: $l_j\in [0.05, 0.15]$, $u_j\in [0.55, 0.65]$ for $j = 1,\ldots,4$, and $k_1\in [0.3, 0.7]$.

To analyse the variability of the results, three different measures were used. These included the mean absolute percentage difference, which was used to measure the difference in uwTOPSIS scores, and the Kendall–tau distance and average ranking position difference, which were used to measure the difference between the rankings produced by uwTOPSIS.

\begin{figure}[ht]
    \centering
    \includegraphics[width =.85\linewidth]{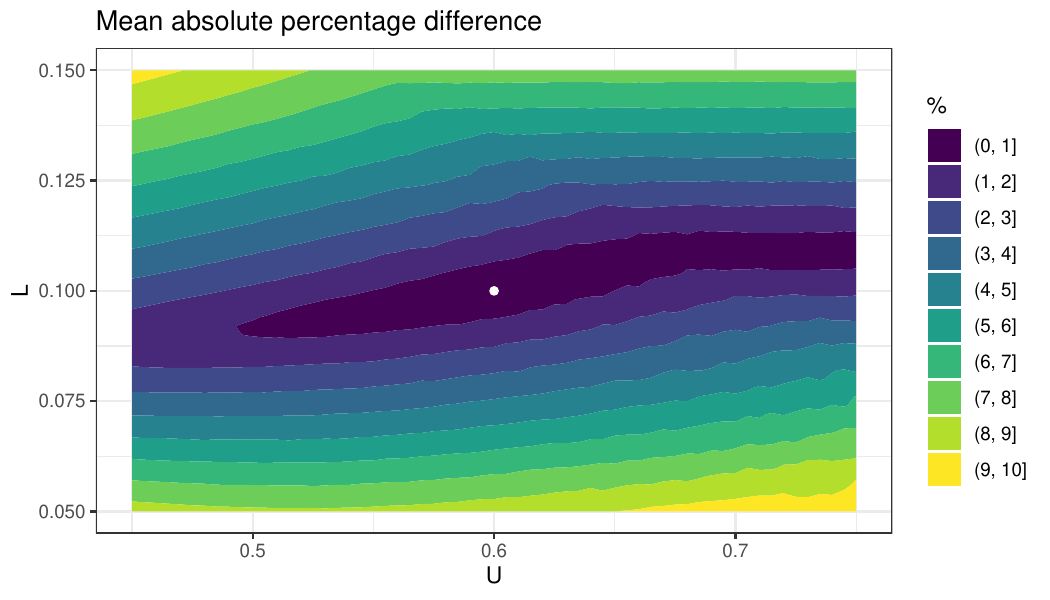}
    \caption{Sensitivity analysis of the uwTOPSIS score concerning the weight bounds. The color scale denotes the values of the mean absolute percentage difference. The white dot marks the value of the lower and upper limits used in the study (L = 0.1, U = 0.6).}
    \label{fig:MAPD}
\end{figure}

\begin{figure}[ht]
    \centering
    \includegraphics[width =\linewidth]{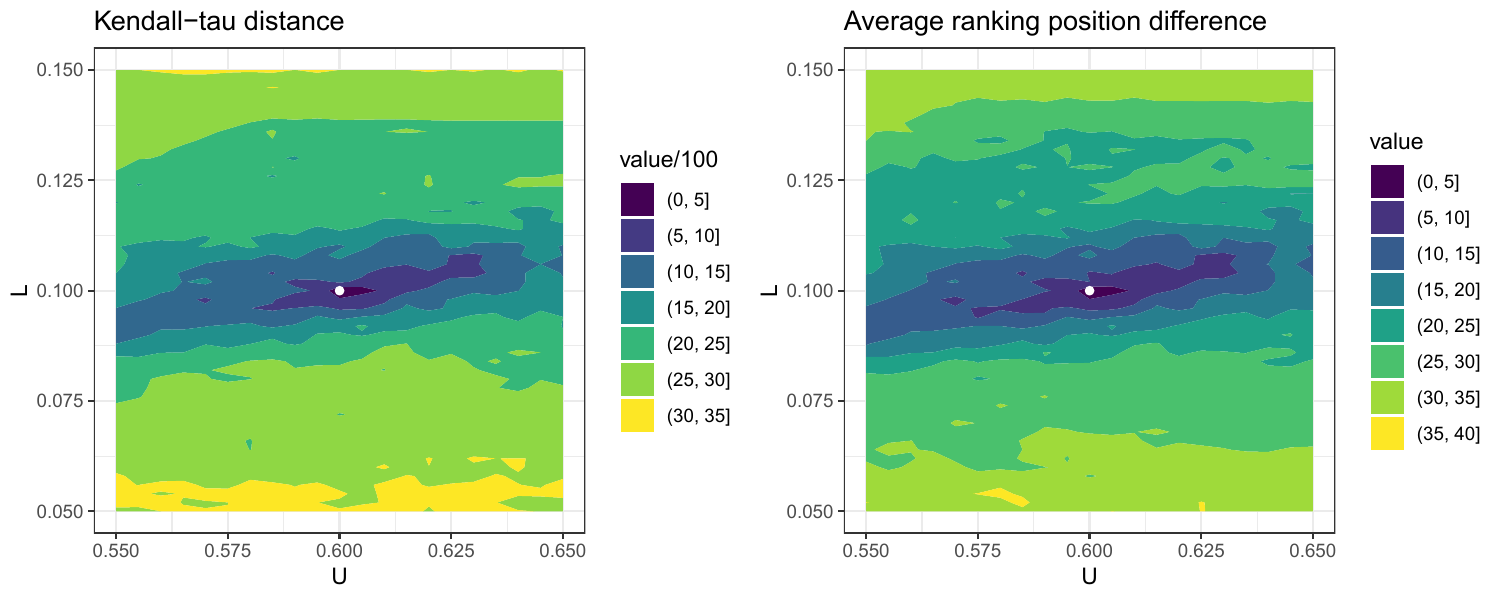}
    \caption{Sensitivity analysis of the uwTOPSIS ranking concerning the weight bounds. The color scale denotes the values of the Kendall-tau distance (left figure) and the average ranking position difference (right figure). The white dot marks the value of the lower and upper limits used in the study (L = 0.1, U = 0.6).}
    \label{fig:Mdif}
\end{figure}

The results of the sensitivity analysis with respect to the lower and upper limits of the weights are presented in Figures \ref{fig:MAPD} and \ref{fig:Mdif}. Figure \ref{fig:MAPD} shows the impact of varying the limits on the uwTOPSIS score. The range of variation of the upper limit ($u$) was slightly extended to enhance the graphical representation. In the most unfavourable scenario, the score variation was approximately 10\%. Figure \ref{fig:Mdif} shows the difference in rankings. Both the Kendall–tau distance and the average ranking position difference revealed comparable patterns, albeit with different scales.

\begin{figure}[ht]
    \centering
    \includegraphics[width =0.7\linewidth]{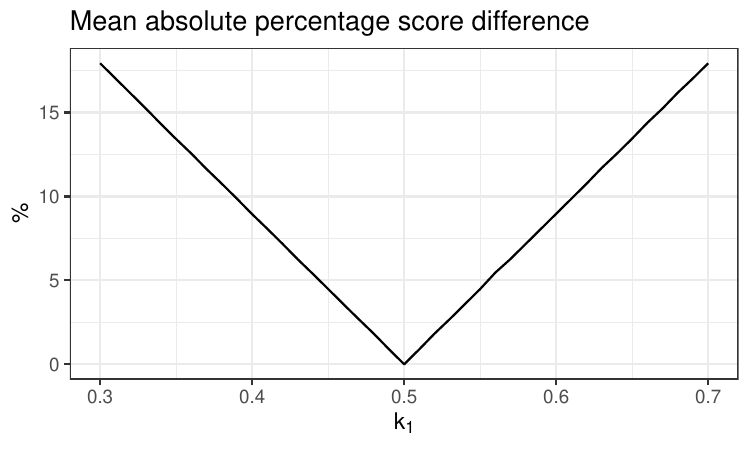}
    \caption{Mean absolute percentage difference of the uwTOPSIS score for the $k_1$ parameter. The value used in the study ($k_1 = 0.5$) corresponds to the point where the MAPD vanishes.}
    \label{fig:k1_MAPSD}
\end{figure}

\begin{figure}[ht]
    \centering
    \includegraphics[width =\linewidth]{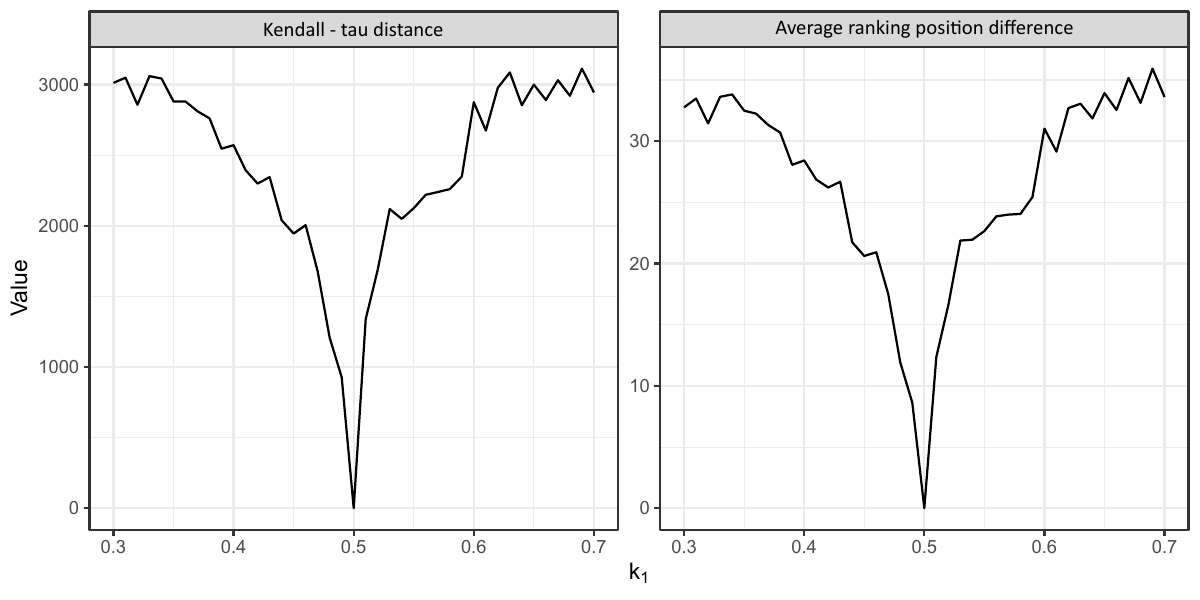}
    \caption{Sensitivity analysis of the uwTOPSIS ranking for the $k_1$ parameter. The curve represents the values of the Kendall-tau distance (left figure) and the average ranking position difference (right figure).}
    \label{fig:k1_diffpos}
\end{figure}

With regard to the sensitivity analysis of the $k_1$ parameter, the corresponding results are shown in Figures \ref{fig:k1_MAPSD} and \ref{fig:k1_diffpos}. The outcomes obtained are analogous to those obtained for the limits of the weights. Sensitivity analysis demonstrated the transparency and robustness of the proposed methodology, because the changes observed in both the uwTOPSIS score and ranking positions were consistent with the modifications made to the parameters.

\section{Conclusions}\label{sec:conclusions}

In this study, we proposed a methodology to analyse the effect of the performance of a certain activity (training courses) on a population. The proposed methodology is completely data-driven (contrary to many studies on the same topic, which are based on opinion surveys), and bottom-up. In particular, the proposed methodology starts by analysing the performance of each participant who took a course. Then, these performances can be aggregated to define the performance of the given course. Finally, courses are ordered using the uwTOPSIS method.

The control group was selected based on the definition of the WLC, which parameterises the working life of individuals and allows comparisons with other subjects. In this way, by means of a clustering process, we can define the control group with which to compare the results of each participant who took each course. This novel methodology permits the definition of control groups in experiments that use pre-existing databases, which can be expanded over a long period of time with heterogeneous individuals. Nevertheless, it requires access to a substantial quantity of data from disparate databases that must be cross-referenced. This requires significant collaboration and coordination between the various institutions.

With regard to the proposed multi-criteria method, uwTOPSIS has an advantage over other weighted methods. In the absence of expert information, the decision maker should only provide some rough lower and upper limits for the weights of each criterion to reduce subjectivity during the decision-making process. This innovative multi-criteria method cannot be compared with other known multi-criteria techniques because fixed weights are not assigned a priori to the criteria; instead, the upper and lower bounds are set. The development of other multi-criteria methods with the introduction of weight bounds could prove an interesting avenue for future research, potentially enabling a comparison of the results of this methodology with those of new approaches.

A sensitivity analysis was conducted to analyse the effect of changing the definition of subjective parameters used in uwTOPSIS on the final score and ranking. This analysis demonstrated the transparency and robustness of the proposed methodology because the observed changes were consistent with the modifications made.

Results of the multi-criteria uwTOPSIS method demonstrated that training courses related to the professional families of ADM, HOT and CSS have a positive effect on employability and improve the working conditions of citizens. Nevertheless, the most efficient courses in the ranking were those that were least demanded by citizens. 

Policymakers can optimise the benefits of lifelong learning by strategically addressing challenges and leveraging opportunities presented in this study. By investing in data quality, engaging stakeholders, building technical capacity, ensuring system flexibility, securing funding and integrating existing systems, policymakers can create robust frameworks that enhance the effectiveness of training programmes. These findings may inform policy decisions on the allocation of public resources, the identification of less efficient training courses for discontinuation and the promotion of courses with a high impact on the employability of citizens. This will ultimately result in enhanced educational outcomes, enhanced employability and a more adaptable and proficient workforce.

\bibliography{references}

\begin{thebibliography}{44}
\providecommand{\natexlab}[1]{#1}
\providecommand{\url}[1]{\texttt{#1}}
\expandafter\ifx\csname urlstyle\endcsname\relax
  \providecommand{\doi}[1]{doi: #1}\else
  \providecommand{\doi}{doi: \begingroup \urlstyle{rm}\Url}\fi

\bibitem[Annoni and Kozovska(2010)]{Annoni2010}
P.~Annoni and K.~Kozovska.
\newblock {EU Regional Competitiveness Index (RCI) 2010}.
\newblock Technical Report EUR 24346 EN, Publications Office of the European
  Union, Luxembourg (Luxembourg), May 2010.
\newblock URL
  \url{https://publications.jrc.ec.europa.eu/repository/handle/JRC58169}.

\bibitem[Babos et~al.(2015)Babos, Lubyova, and Studen{\'a}]{Babos2015}
P.~Babos, M.~Lubyova, and I.~Studen{\'a}.
\newblock Lifelong learning is a growing factor in employability. policy brief,
  proceedings of lllight’in’europe research project, 2015.
\newblock URL
  \url{"https://www.lllightineurope.com/fileadmin/lllightineurope/download/LLLight\_LLL\_is\_growing\_factor\_in\_employability\_policybrief\_C7\_20150922.pdf"}.
\newblock Accesed 2023/09/21.

\bibitem[Benítez and Liern(2021)]{Benitez2021}
R.~Benítez and V.~Liern.
\newblock Unweighted topsis: a new multi-criteria tool for sustainability
  analysis.
\newblock \emph{International Journal of Sustainable Development \& World
  Ecology}, 28\penalty0 (1):\penalty0 36--48, 2021.
\newblock \doi{10.1080/13504509.2020.1778583}.
\newblock URL \url{https://doi.org/10.1080/13504509.2020.1778583}.

\bibitem[Blasco-Blasco et~al.(2021{\natexlab{a}})Blasco-Blasco,
  Liern-Garc{\'i}a, L{\'o}pez-Garc{\'i}a, and Parada-Rico]{Blasco-Blasco2021}
O.~Blasco-Blasco, M.~Liern-Garc{\'i}a, A.~L{\'o}pez-Garc{\'i}a, and S.~E.
  Parada-Rico.
\newblock An academic performance indicator using flexible multi-criteria
  methods.
\newblock \emph{Mathematics}, 9\penalty0 (19), 2021{\natexlab{a}}.
\newblock ISSN 2227-7390.
\newblock \doi{10.3390/math9192396}.
\newblock URL \url{https://doi.org/10.3390/math9192396}.

\bibitem[Blasco-Blasco et~al.(2021{\natexlab{b}})Blasco-Blasco, Liern-García,
  López-García, and Parada-Rico]{Blasco}
O.~Blasco-Blasco, M.~Liern-García, A.~López-García, and S.~E. Parada-Rico.
\newblock An academic performance indicator using flexible multi-criteria
  methods.
\newblock \emph{Mathematics}, 9:\penalty0 1--19, 2021{\natexlab{b}}.
\newblock ISSN 2227-7390.
\newblock \doi{https://doi.org/10.3390/math9192396}.

\bibitem[Cables et~al.(2016)Cables, Lamata, and Verdegay]{Cables}
E.~Cables, M.~Lamata, and J.~Verdegay.
\newblock {RIM-reference ideal method in multicriteria decision making}.
\newblock \emph{Information Sciences}, 337-338:\penalty0 1 -- 10, 2016.
\newblock ISSN 0020-0255.
\newblock \doi{https://doi.org/10.1016/j.ins.2015.12.011}.
\newblock URL \url{https://doi.org/10.1016/j.ins.2015.12.011}.

\bibitem[Can{\'o}s and Liern(2008)]{Canos}
L.~Can{\'o}s and V.~Liern.
\newblock Soft computing-based aggregation methods for human resource
  management.
\newblock \emph{European Journal of Operational Research}, 189\penalty0
  (3):\penalty0 669 -- 681, 2008.
\newblock ISSN 0377-2217.
\newblock \doi{https://doi.org/10.1016/j.ejor.2006.01.054}.
\newblock URL \url{https://doi.org/10.1016/j.ejor.2006.01.054}.

\bibitem[Conejero et~al.(2021{\natexlab{a}})Conejero, Preciado, Prieto, Bas,
  and Bolós]{conejero2021b}
J.~Conejero, J.~Preciado, A.~Prieto, M.~Bas, and V.~Bolós.
\newblock Applying data driven decision making to rank vocational and
  educational training programs with topsis.
\newblock \emph{Decision Support Systems}, 142:\penalty0 113470,
  2021{\natexlab{a}}.
\newblock ISSN 0167-9236.
\newblock \doi{https://doi.org/10.1016/j.dss.2020.113470}.
\newblock URL
  \url{https://www.sciencedirect.com/science/article/pii/S0167923620302256}.

\bibitem[Conejero et~al.(2021{\natexlab{b}})Conejero, Preciado,
  Fernández-García, Prieto, and Rodríguez-Echeverría]{conejero2021a}
J.~M. Conejero, J.~C. Preciado, A.~J. Fernández-García, A.~E. Prieto, and
  R.~Rodríguez-Echeverría.
\newblock Towards the use of data engineering, advanced visualization
  techniques and association rules to support knowledge discovery for public
  policies.
\newblock \emph{Expert Systems with Applications}, 170:\penalty0 114509,
  2021{\natexlab{b}}.
\newblock ISSN 0957-4174.
\newblock \doi{https://doi.org/10.1016/j.eswa.2020.114509}.
\newblock URL
  \url{https://www.sciencedirect.com/science/article/pii/S0957417420311532}.

\bibitem[Elena(2014)]{Elena2014}
R.~Elena.
\newblock Effectiveness evaluation of training programmes for disadvantaged
  targets.
\newblock \emph{Procedia - Social and Behavioral Sciences}, 141:\penalty0
  1239--1243, 2014.
\newblock ISSN 1877-0428.
\newblock \doi{https://doi.org/10.1016/j.sbspro.2014.05.213}.
\newblock URL
  \url{https://www.sciencedirect.com/science/article/pii/S1877042814036374}.
\newblock 4th World Conference on Learning Teaching and Educational Leadership
  (WCLTA-2013).

\bibitem[{European Comission}(2001)]{informeCE}
{European Comission}.
\newblock Making a european area of lifelong learning a reality. communication
  from the commission. com (2001) 678 final, 2001.
\newblock URL \url{http://aei.pitt.edu/42878/}.

\bibitem[Farjad(2012)]{Farjad2012}
S.~Farjad.
\newblock {The Evaluation Effectiveness of Training Courses in University by
  Kirkpatrick Model (Case Study: Islamshahr University)}.
\newblock \emph{Procedia - Social and Behavioral Sciences}, 46:\penalty0
  2837--2841, 2012.
\newblock ISSN 18770428.
\newblock \doi{10.1016/j.sbspro.2012.05.573}.
\newblock URL \url{http://dx.doi.org/10.1016/j.sbspro.2012.05.573}.

\bibitem[Gal et~al.(2007)Gal, Janowszky, and Juhasz-Fodor]{Gal2007}
J.~Gal, S.~Janowszky, and T.~Juhasz-Fodor.
\newblock Lifelong learning in the hungarian education system.
\newblock In J.~Gal, editor, \emph{MERLIN– A t{\"o}rt{\'e}netmes{\'e}l{\'e}s,
  mint az {\'e}lethosszig tart{\'o}tanul{\'a}s egyik form{\'a}ja}, pages
  125--130. IM Informatikai Magániskola Kft, 2007.
\newblock ISBN 978-963-06-3267-6.

\bibitem[Gil-Aluja(1999)]{Gil}
J.~Gil-Aluja.
\newblock \emph{Elements for a Theory of Decision in Uncertainty}.
\newblock Applied Optimization 32. Springer US, 1 edition, 1999.
\newblock ISBN 978-1-4419-4817-5,978-1-4757-3011-1.

\bibitem[Hwang and Yoon(1981)]{Hwang1981}
C.-L. Hwang and K.~Yoon.
\newblock \emph{Multiple Attribute Decision Making}, volume 186.
\newblock Springer Berlin Heidelberg, 1981.
\newblock ISBN 978-3-540-10558-9.
\newblock \doi{10.1007/978-3-642-48318-9}.
\newblock URL \url{http://link.springer.com/10.1007/978-3-642-48318-9}.

\bibitem[J{\"a}rlstr{\"o}m et~al.(2020)J{\"a}rlstr{\"o}m, Brandt, and
  Rajala]{Jarlstrom2020}
M.~J{\"a}rlstr{\"o}m, T.~Brandt, and A.~Rajala.
\newblock The relationship between career capital and career success among
  finnish knowledge workers.
\newblock \emph{Baltic Journal of Management}, 15\penalty0 (5):\penalty0
  687--706, Jan 2020.
\newblock ISSN 1746-5265.
\newblock \doi{10.1108/BJM-10-2019-0357}.
\newblock URL \url{https://doi.org/10.1108/BJM-10-2019-0357}.

\bibitem[Johnson(2022)]{nlopt}
S.~G. Johnson.
\newblock \emph{The NLopt nonlinear-optimization package}, 2022.
\newblock URL \url{{http://github.com/stevengj/nlopt}}.
\newblock Accessed September 30, 2022.

\bibitem[Kirkpatrick(2006)]{Kirkpatrick2006}
D.~L. Kirkpatrick.
\newblock \emph{Evaluating Training Programs: The Four Levels}.
\newblock Berrett-Koehler Publishers, 3rd edition, 2006.
\newblock ISBN 1576753484.

\bibitem[Leamer(1983)]{Leamer1983}
E.~E. Leamer.
\newblock Let's take the con out of econometrics.
\newblock \emph{The American Economic Review}, 73\penalty0 (1):\penalty0
  31--43, 1983.
\newblock ISSN 00028282.
\newblock URL \url{http://www.jstor.org/stable/1803924}.

\bibitem[Liern and P{\'e}rez-Gladish(2022)]{Liern2022}
V.~Liern and B.~P{\'e}rez-Gladish.
\newblock Multiple criteria ranking method based on functional proximity index:
  un-weighted topsis.
\newblock \emph{Annals of Operations Research}, 311\penalty0 (2):\penalty0
  1099--1121, Apr 2022.
\newblock ISSN 1572-9338.
\newblock \doi{10.1007/s10479-020-03718-1}.
\newblock URL \url{https://doi.org/10.1007/s10479-020-03718-1}.

\bibitem[Liern et~al.(2021)Liern, P{\'e}rez-Gladish, Rubiera-Moroll{\'o}n, and
  M'Zali]{Liern2021}
V.~Liern, B.~P{\'e}rez-Gladish, F.~Rubiera-Moroll{\'o}n, and B.~M'Zali.
\newblock Residential choice from a multiple criteria sustainable perspective.
\newblock \emph{Annals of Operations Research}, Dec 2021.
\newblock ISSN 1572-9338.
\newblock \doi{10.1007/s10479-021-04480-8}.
\newblock URL \url{https://doi.org/10.1007/s10479-021-04480-8}.

\bibitem[L{\'o}pez-Garc{\'i}a et~al.(2023)L{\'o}pez-Garc{\'i}a, Liern, and
  P{\'e}rez-Gladish]{Lopez-Garcia2023}
A.~L{\'o}pez-Garc{\'i}a, V.~Liern, and B.~P{\'e}rez-Gladish.
\newblock Determining the underlying role of corporate sustainability criteria
  in a ranking problem using uw-topsis.
\newblock \emph{Annals of Operations Research}, 2023.
\newblock ISSN 1572-9338.
\newblock \doi{10.1007/s10479-023-05543-8}.
\newblock URL \url{https://doi.org/10.1007/s10479-023-05543-8}.

\bibitem[López-García et~al.(2023)López-García, Blasco-Blasco,
  Liern-García, and Parada-Rico]{Lopez-Garcia}
A.~López-García, O.~Blasco-Blasco, M.~Liern-García, and S.~E. Parada-Rico.
\newblock Early detection of students’ failure using machine learning
  techniques.
\newblock \emph{Operations Research Perspectives}, 11:\penalty0 1--11, 2023.
\newblock ISSN 2214-7160.
\newblock \doi{https://doi.org/10.1016/j.orp.2023.100292}.

\bibitem[Ouenniche et~al.(2018)Ouenniche, Pérez-Gladish, and Bouslah]{Jamal}
J.~Ouenniche, B.~Pérez-Gladish, and K.~Bouslah.
\newblock An out-of-sample framework for topsis-based classifiers with
  application in bankruptcy prediction.
\newblock \emph{Technological Forecasting and Social Change}, 131:\penalty0 111
  -- 116, 2018.
\newblock ISSN 0040-1625.
\newblock \doi{https://doi.org/10.1016/j.techfore.2017.05.034}.

\bibitem[Pamucar et~al.(2018)Pamucar, Stevic, and Sremac]{Pamucar2018}
D.~Pamucar, Z.~Stevic, and S.~Sremac.
\newblock A new model for determining weight coefficients of criteria in mcdm
  models: Full consistency method (fucom).
\newblock \emph{Symmetry}, 10\penalty0 (9), 2018.
\newblock ISSN 2073-8994.
\newblock \doi{10.3390/sym10090393}.
\newblock URL \url{https://www.mdpi.com/2073-8994/10/9/393}.

\bibitem[Park and Jun(2009)]{Park2009}
H.-S. Park and C.-H. Jun.
\newblock A simple and fast algorithm for k-medoids clustering.
\newblock \emph{Expert Systems with Applications}, 36\penalty0 (2, Part
  2):\penalty0 3336--3341, 2009.
\newblock ISSN 0957-4174.
\newblock \doi{https://doi.org/10.1016/j.eswa.2008.01.039}.
\newblock URL
  \url{https://www.sciencedirect.com/science/article/pii/S095741740800081X}.

\bibitem[P{\'e}rez-Gladish et~al.(2021)P{\'e}rez-Gladish, Ferreira, and
  Zopounidis]{Perez-Gladish2021}
B.~P{\'e}rez-Gladish, F.~A.~F. Ferreira, and C.~Zopounidis.
\newblock Mcdm/a studies for economic development, social cohesion and
  environmental sustainability: introduction.
\newblock \emph{International Journal of Sustainable Development {\&} World
  Ecology}, 28\penalty0 (1):\penalty0 1--3, Jan 2021.
\newblock ISSN 1350-4509.
\newblock \doi{10.1080/13504509.2020.1821257}.
\newblock URL \url{https://doi.org/10.1080/13504509.2020.1821257}.

\bibitem[Powell(1994)]{Powell1994}
M.~Powell.
\newblock \emph{Advances in Optimization and Numerical Analysis}, volume
  10.1007/978-94-015-8330-5, chapter {A Direct Search Optimization Method That
  Models the Objective and Constraint Functions by Linear Interpolation}, pages
  51--67.
\newblock Susana Gomez and Jean-Pierre Hennart, 1994.
\newblock ISBN 978-90-481-4358-0,978-94-015-8330-5.
\newblock \doi{10.1007/978-94-015-8330-5\_4}.
\newblock URL \url{http://dx.doi.org/10.1007/978-94-015-8330-5\_4}.

\bibitem[{R Core Team}(2019)]{Rcore}
{R Core Team}.
\newblock \emph{R: A Language and Environment for Statistical Computing}.
\newblock R Foundation for Statistical Computing, Vienna, Austria, 2019.
\newblock URL \url{https://www.R-project.org/}.

\bibitem[Ram{\'{\i}}k and {\'{\i}}m{\'{a}}nek(1985)]{Ramik}
J.~Ram{\'{\i}}k and J.~{\'{\i}}m{\'{a}}nek.
\newblock {Inequality relation between fuzzy numbers and its use in fuzzy
  optimization}.
\newblock \emph{Fuzzy Sets and Systems}, 16\penalty0 (2):\penalty0 123--138,
  jul 1985.
\newblock \doi{10.1016/s0165-0114(85)80013-0}.

\bibitem[Rotar(2021)]{Rotar2021}
L.~J. Rotar.
\newblock {Evaluation of the effectiveness of employment programmes on young
  unemployed people}.
\newblock \emph{Engineering Economics}, 32\penalty0 (1):\penalty0 60--69, 2021.
\newblock ISSN 20295839.
\newblock \doi{10.5755/j01.ee.32.1.23276}.

\bibitem[Saaty and Vargas(2012)]{Saaty2012}
T.~Saaty and L.~Vargas.
\newblock \emph{Models, methods, concepts \& applications of the analytic
  hierarchy process}.
\newblock Springer, 2012.
\newblock ISBN 9781461435969.
\newblock \doi{10.1007/978-1-4614-3597-6}.
\newblock URL
  \url{http://link.springer.com/content/pdf/10.1007/978-1-4614-3597-6.pdf}.

\bibitem[Sackman(1974)]{Sackman1974}
H.~Sackman.
\newblock \emph{Delphi Assessment: Expert Opinion, Forecasting, and Group
  Process}.
\newblock RAND Corporation, Santa Monica, CA, 1974.

\bibitem[Sanulita et~al.(2024)Sanulita, Hendriyanto, Lestari, Ramli, and
  Arifudin]{Sanulita}
H.~Sanulita, D.~Hendriyanto, N.~C. Lestari, A.~Ramli, and O.~Arifudin.
\newblock Analysis of the effectiveness of audio visual learning media based on
  macromedia flash usage on school program of increasing student learning
  motivation.
\newblock \emph{Journal on Education}, 6:\penalty0 12641--12650, 2024.
\newblock ISSN 2654-5497.
\newblock \doi{https://doi.org/10.31004/joe.v6i2.5121}.

\bibitem[Schubert and Rousseeuw(2019)]{Sch2019}
E.~Schubert and P.~J. Rousseeuw.
\newblock Faster k-medoids clustering: Improving the pam, clara, and clarans
  algorithms.
\newblock In G.~Amato, C.~Gennaro, V.~Oria, and M.~Radovanovi{\'{c}}, editors,
  \emph{Similarity Search and Applications}, pages 171--187, Cham, 2019.
  Springer International Publishing.
\newblock ISBN 978-3-030-32047-8.

\bibitem[Shannon(1948)]{Shannon1948}
C.~E. Shannon.
\newblock A mathematical theory of communication.
\newblock \emph{The Bell System Technical Journal}, 27\penalty0 (3):\penalty0
  379--423, 1948.
\newblock \doi{10.1002/j.1538-7305.1948.tb01338.x}.

\bibitem[Sharma et~al.(2024)Sharma, Jain, Mogaji, and Babbilid]{Sharma}
H.~Sharma, V.~Jain, E.~Mogaji, and A.~S. Babbilid.
\newblock Blended learning and augmented employability: a multi-stakeholder
  perspective of the microcredentialing ecosystem in higher education.
\newblock \emph{International Journal of Educational Management}, 38, 2024.
\newblock ISSN 0951-354X.
\newblock \doi{10.1108/ijem-12-2022-0497}.

\bibitem[Srinivasan and Shocker(1973)]{Srinivasan1973}
V.~Srinivasan and A.~D. Shocker.
\newblock Linear programming techniques for multidimensional analysis of
  preferences.
\newblock \emph{Psychometrika}, 38\penalty0 (3):\penalty0 337--369, Sep 1973.
\newblock ISSN 1860-0980.
\newblock \doi{10.1007/BF02291658}.
\newblock URL \url{https://doi.org/10.1007/BF02291658}.

\bibitem[Tibshirani et~al.(2001)Tibshirani, Walther, and Hastie]{Tib2001}
R.~Tibshirani, G.~Walther, and T.~Hastie.
\newblock Estimating the number of clusters in a data set via the gap
  statistic.
\newblock \emph{Journal of the Royal Statistical Society: Series B (Statistical
  Methodology)}, 63\penalty0 (2):\penalty0 411--423, 2001.
\newblock \doi{10.1111/1467-9868.00293}.
\newblock URL
  \url{https://rss.onlinelibrary.wiley.com/doi/abs/10.1111/1467-9868.00293}.

\bibitem[Towler et~al.(2019)Towler, Kemp, {Mike Burton}, Dunn, Wayne, Moreton,
  and White]{Towler2019}
A.~Towler, R.~I. Kemp, A.~{Mike Burton}, J.~D. Dunn, T.~Wayne, R.~Moreton, and
  D.~White.
\newblock {Do professional facial image comparison training courses work?}
\newblock \emph{PLoS ONE}, 14\penalty0 (2):\penalty0 1--17, 2019.
\newblock ISSN 19326203.
\newblock \doi{10.1371/journal.pone.0211037}.

\bibitem[Tzeng and Huang(2011)]{Haang2011}
G.-H. Tzeng and J.-J. Huang.
\newblock \emph{Multiple Attribute Decision Making: Methods and Applications}.
\newblock Chapman and Hall /CRC, 1 edition, 2011.
\newblock ISBN 1439861579,9781439861578.

\bibitem[Vinogradova et~al.(2018)Vinogradova, Podvezko, and
  Zavadskas]{Vinogradova2018}
I.~Vinogradova, V.~Podvezko, and E.~K. Zavadskas.
\newblock The recalculation of the weights of criteria in mcdm methods using
  the bayes approach.
\newblock \emph{Symmetry}, 10\penalty0 (6), 2018.
\newblock ISSN 2073-8994.
\newblock \doi{10.3390/sym10060205}.
\newblock URL \url{https://www.mdpi.com/2073-8994/10/6/205}.

\bibitem[{Ypma, J.}(2013)]{nloptr}
{Ypma, J.}
\newblock \emph{Introduction to NLoptr: An R interface to NLopt.}, 2013.
\newblock URL
  \url{{http://cran.r-project.org/web/packages/nloptr/vignettes/nloptr.pdf}}.
\newblock Accessed December 23, 2019.

\bibitem[Zavadskas and Podvezko(2016)]{Zavadskas2016}
E.~K. Zavadskas and V.~Podvezko.
\newblock Integrated determination of objective criteria weights in mcdm.
\newblock \emph{International Journal of Information Technology \& Decision
  Making}, 15\penalty0 (02):\penalty0 267--283, 2016.
\newblock \doi{10.1142/S0219622016500036}.
\newblock URL \url{https://doi.org/10.1142/S0219622016500036}.

\end{thebibliography}
\bibliographystyle{abbrvnat}

\end{document}